\documentclass[twocolumn]{jpsj3} 
\usepackage{color}
\title{
Magnetization Process 
of the Spin-1/2 Triangular-Lattice Heisenberg Antiferromagnet 
with Next-Nearest-Neighbor Interactions 
\\
\-- 
Plateau or Nonplateau 
\--
}
\catcode`\@=11
\def\simle{\mathrel{\mathpalette\@versim<}}   
\def\simge{\mathrel{\mathpalette\@versim>}}   
\def\@versim#1#2{\lower2.5pt\vbox{\baselineskip0pt \lineskip-.5pt
   \ialign{$\m@th#1\hfil##\hfil$\crcr#2\crcr\sim\crcr}}}
\catcode`\@=12

\author{Hiroki Nakano$^{1}$
\thanks{E-mail: hnakano@sci.u-hyogo.ac.jp} 
and 
T\^oru Sakai$^{1,2}$
\thanks{E-mail: sakai@spring8.or.jp}
}

\inst{$^{1}$Graduate School of Material Science, 
University of Hyogo,
Kamigori, 
Hyogo 678-1297, Japan \\
$^{2}$
National Institutes for Quantum and Radiological Science and Technology 
(QST),
SPring-8
Sayo, Hyogo 679-5148, Japan 
}

\recdate{\today}

\abst{
An $S=1/2$ triangular-lattice Heisenberg antiferromagnet 
with next-nearest-neighbor interactions 
is investigated under a magnetic field 
by the numerical-diagonalization method. 
It is known that, 
in both cases of weak and strong next-nearest-neighbor interactions, 
this system reveals a magnetization plateau 
at one-third of the saturated magnetization. 
We examine the stability of this magnetization plateau
when the amplitude of next-nearest-neighbor interactions 
is varied. 
We find that a nonplateau region appears between the plateau phases 
in the cases of weak and strong next-nearest-neighbor interactions. 
}

\begin{document}
\maketitle

\section{Introduction} 

Frustration has attracted the attention 
of many researchers in condensed-matter physics 
because frustration becomes a source of nontrivial quantum phenomena 
in various systems of physics. 
In magnetic materials, frustrations occur, for example, 
when antiferromagnetic interactions of a system form a triangle. 
The triangular-lattice antiferromagnet is a typical case. 
Since Anderson\cite{Anderson_tri} pointed out that this model is 
a possible candidate for the realization 
of a spin-liquid ground state 
owing to frustrations in the system,
the $S=1/2$ Heisenberg model in particular has been 
studied\cite{Huse_Elser,Jolicour_LGuillou,Bernu1992,
Cubukov1992,Bernu1994,Singh_Huse,Lecheminant1995,
Leung_Runge,Richter_lecture2004,Sakai_HN_PRBR,DYamamoto2013,
Weihong1999,Yunoki2006,Starykh2007,Heidarian2009,Reuther2011,
Weichselbaum2011,Ghamari2011,Harada2011,s1tri_LRO,Starykh_review2015}. 
On the basis of extensive studies of the ground state of this system, 
many physicists believe that the symmetry-breaking state 
with the so-called 120-degree structure is realized 
in the ground state.  
On the other hand, a recent large-scale numerical study\cite{Kulagin_PRL2013} 
suggested the absence of such breaking;  
this issue remains controversial even now. 

One nontrivial phenomenon in 
the quantum Heisenberg antiferromagnet 
on the triangular lattice under magnetic fields 
is the magnetization plateau. 
This phenomenon appears at one-third of the saturated magnetization 
in the magnetization curve.  
It is known that the plateau on the triangular-lattice antiferromagnet 
appears by the so-called {\it order-by-disorder} mechanism 
at zero temperature 
even though the corresponding classical system does not show 
the plateau\cite{Alicea_PRL2009}. 
In an early stage of investigations of the plateau, 
the presence of the plateau was just a theoretical prediction. 
However, experimental realizations have been reported, 
for examples, for Ba$_{3}$CoSb$_{2}$O$_{9}$ 
in the $S=1/2$ case\cite{Shirata_Shalf_PRL2012} 
and for Ba$_{3}$NiSb$_{2}$O$_{9}$ in the $S=1$ case\cite{Shirata_S1tri_2011}. 
On the other hand, 
the destabilization of this plateau was studied 
from the viewpoint of the effect of randomness in the system
\cite{KWatanabe_JPSJ2015,Shimokawa_JPSJ2015}. 

Under the above circumstances, we are faced with a question: 
What else can destabilize the plateau? 
A possible candidate is additional interactions 
at a next-nearest-neighbor (NNN) pair. 
NNN interactions in the $S=1/2$ triangular-lattice 
Heisenberg antiferromagnet have already been 
studied\cite{Bernu1992,Cubukov1992,Lecheminant1995,Wietek_arXiv2016,
Bieri_PRL}; 
however, the studies were carried out for the system 
without an external magnetic field. 
The purpose of the present study is 
to clarify how the plateau behaves in the presence of NNN interactions 
at zero temperature. 
In particular, 
we focus our attention on whether or not a plateau of this height
is present during the variation of NNN interactions. 
Our numerical-diagonalization study provides us with 
information on a new phase transition driven by NNN interactions. 

This paper is organized as follows. 
In the next section, the model studied here is introduced. 
The method is also explained. 
The third section is devoted 
to the presentation and discussion of our results. 
We first observe magnetization processes
for various amplitudes of NNN interactions. 
In order to understand the behavior observed 
in magnetization processes, 
several analyses are carried out. 
In the final section, we present our conclusion.

\section{Model Hamiltonian and Method} 

\begin{figure}[tb]
\begin{center}
\includegraphics[width=8cm]{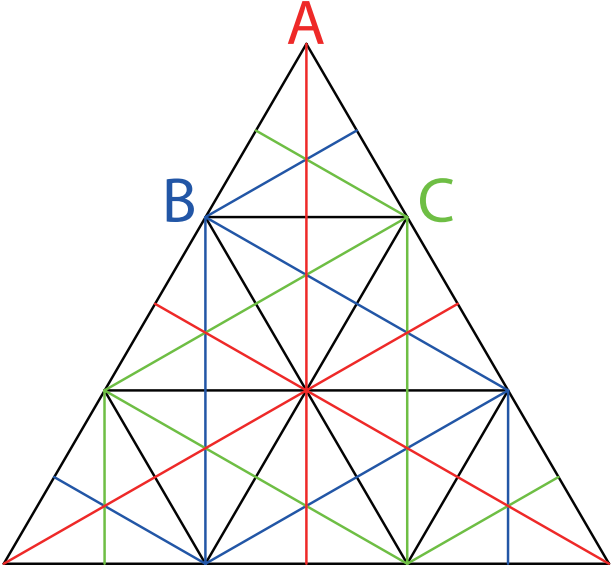}
\end{center}
\caption{
(Color) 
Interaction bonds of triangular-lattice antiferromagnet. 
Black lines denote nearest-neighbor interactions, while 
colored lines denote next-nearest-neighbor interactions. 
Sublattices A, B, and C are explicitly presented. 
}
\label{fig1}
\end{figure}

The Hamiltonian studied here is given by 
${\cal H}={\cal H}_{0} +{\cal H}_{\rm Zeeman}  $, where
\begin{equation}
{\cal H}_{0} 
=
\sum_{(i ,j): \ {\rm n.n.}} J_{1} 
\mbox{\boldmath $S$}_{i}\cdot\mbox{\boldmath $S$}_{j} 
+\sum_{(i,j): \ {\rm n.n.n.}} J_{2} 
\mbox{\boldmath $S$}_{i}\cdot\mbox{\boldmath $S$}_{j} 
, 
\label{Hamiltonian}
\end{equation}
and the Zeeman term is given by 
\begin{equation}
{\cal H}_{\rm Zeeman} = -h \sum_{j} S_{j}^{z} . 
\label{Zeeman}
\end{equation}
Here, $\mbox{\boldmath $S$}_{i}$ 
denotes the $S=1/2$ spin operator at site $i$. 
In this study, we consider 
the case of an isotropic interaction in spin space. 
Site $i$ is assumed to be the vertices 
of a triangular lattice composed of 
bonds of nearest-neighbor interactions $J_{1}$ 
illustrated by the black lines in Fig.~\ref{fig1}. 
The number of spin sites is denoted by $N_{\rm s}$.  
The vertices of the triangular lattice 
are divided into three equivalent sublattices: A, B, and C. 
Note here that each sublattice also forms a triangular lattice 
illustrated by colored lines in Fig.~\ref{fig1}. 
The colored lines represent NNN interactions 
whose amplitudes are all given by $J_{2}$. 
We denote the ratio $J_{2}/J_{1}$ by $r$.  
We consider that all the interactions are antiferromagnetic, 
namely, $J_{1} > 0$ and $J_{2} > 0$. 
Energies are measured in units of $J_{1}$; 
hereafter, we set $J_{1}=1$ 
and examine the case of $r \ge 0$. 
Note here that for $J_{2}=0$, namely, $r=0$, 
the present lattice is identical to the triangular lattice 
without NNN interactions 
and that 
for infinitely large $J_{2}$, namely, $r\rightarrow\infty$, 
the system reduces to three isolated triangular-lattice antiferromagnets. 

\begin{figure}[tb]
\begin{center}
\includegraphics[width=8cm]{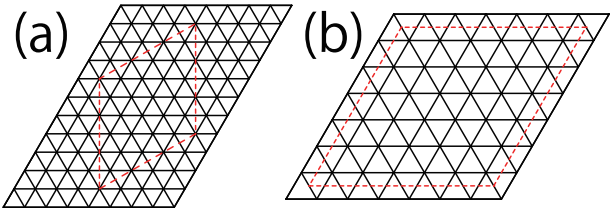}
\end{center}
\caption{
(Color) 
Finite-size clusters investigated here. 
Panels (a) and (b) illustrate the cases for $N_{\rm s}=27$ and 36, 
respectively. Bonds of the next-nearest-neighbor interactions 
are not illustrated. 
}
\label{fig2}
\end{figure}

The finite-size clusters that we treat in the present study 
are depicted in Fig.~\ref{fig2}.  
We examine the cases of $N_{\rm s}=$27 and 36 
under the periodic boundary condition.  
In order to detect the magnetization plateau 
at one-third of the saturated magnetization 
in both limiting cases 
for a vanishing $J_{2}$ and an infinitely large $J_{2}$,  
it is necessary for $N_{\rm s}/9$ to be an integer. 
In order to capture two-dimensionality well, additionally, 
the cluster shapes are assumed to 
be rhombic and to have an inner angle of $\pi/3$. 
The rhombic condition is satisfied 
not only for the triangular lattice of nearest-neighbor interactions 
but also for any of the triangular lattices of each sublattice.  

We calculate the lowest energy of ${\cal H}_{0}$ 
in the subspace belonging to $\sum _j S_j^z=M$ 
by numerical diagonalizations 
based on the Lanczos algorithm and/or Householder algorithm. 
Our diagonalizations are carried out in the basis 
where the $z$-axis 
is taken as the quantized axis of each spin.  
The numerical-diagonalization calculations are unbiased; 
one can therefore obtain reliable information on the system. 
The energy is denoted by $E_{r}(N_{\rm s},M)$, 
where $M$ takes an integer or a half odd integer up 
to the saturation value $M_{\rm sat}$ ($=N_{\rm s}S$) 
for the $N_{\rm s}$-site system with the ratio $r$. 
The normalized magnetization is denoted by $m=M/M_{\rm sat}$. 
We focus our attention on the magnetization process 
at zero temperature. 
For given $N_{\rm s}$ and $r$, we evaluate the magnetic field 
where 
the magnetization increases from $M$ to $M+1$ 
at the field 
\begin{equation}
h=E_{r}(N_{\rm s},M+1) - E_{r}(N_{\rm s},M). 
\label{MH_determining}
\end{equation}
Some of the Lanczos diagonalizations were carried out 
using an MPI-parallelized code that was originally 
developed in the study of Haldane gaps\cite{HN_Terai}. 
The usefulness of our program was confirmed in large-scale 
parallelized calculations\cite{kgm_gap,s1tri_LRO,
HN_TSakai_kgm_1_3,HN_TSakai_kgm_S,HN_YHasegawa_TSakai_dist_shuriken}. 

\section{Results and Discussion} 

\begin{figure}[tb]
\begin{center}
  \includegraphics[width=7cm]{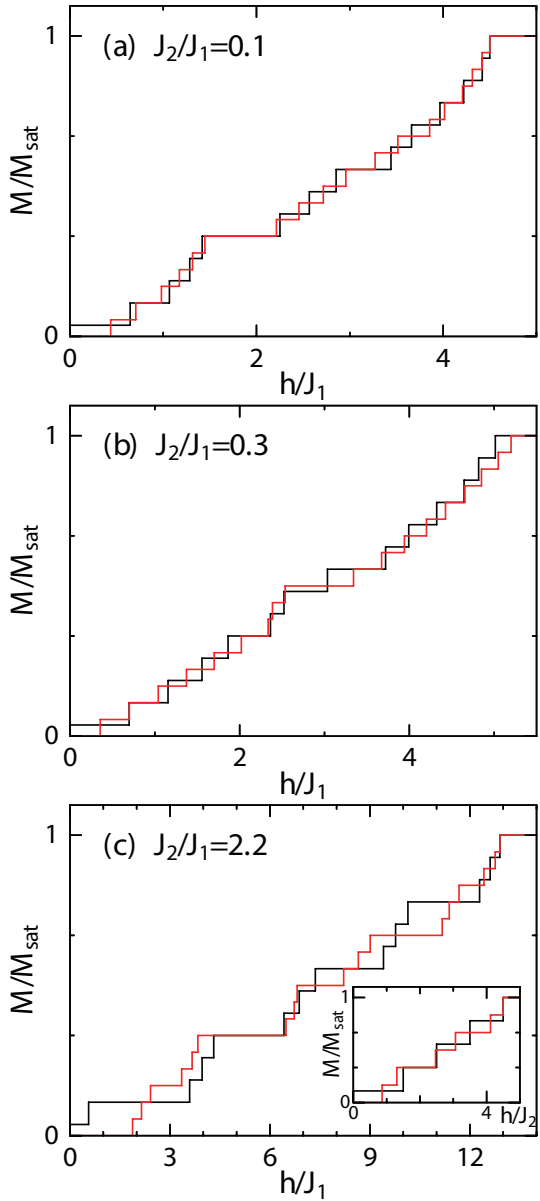}
\end{center}
\caption{
(Color) 
Magnetization curve for various NNN interactions. 
Black and red lines denote the cases for $N_{\rm s}=27$ and 36, respectively. 
The inset of panel (c) depicts the limiting case for $J_{2}\rightarrow\infty$; 
note here that the scale of the abscissa is different 
from that of the main panel. 
}
\label{fig3}
\end{figure}

Now, we depict our numerical results 
of the magnetization process in the three cases 
of $r=J_{2}/J_{1}=0.1$, 0.3, and 2.2 
in Fig.~\ref{fig3}. 
Recall the result in the case of $J_{2}=0$ 
reported in Ref.~\ref{Sakai_HN_PRBR} and 
let us compare the result of $J_{2}=0$ in Ref.~\ref{Sakai_HN_PRBR} 
and the present result in Fig.~\ref{fig3}(a). 
Since NNN interactions are small, 
significant differences are not observed 
between the two cases. 
Note here that the presence of the $m=1/3$ plateau is clearly detected 
even when NNN interactions are switched on 
when the amplitude of the interaction is not large. 
Next, let us observe the case of $r=2.2$ presented in Fig.~\ref{fig3}(c). 
The result for $N_{\rm s}=27$ shows that 
there are some finite-size steps with large widths 
and other steps with significantly smaller widths. 
The large-width steps appear once every three steps. 
This behavior should be compared with 
that in the case of an infinitely large $J_{2}$ 
in the inset of Fig.~\ref{fig3}(c). 
In the limiting case, the system is reduced to three isolated 
triangular-lattice antiferromagnets of $N_{\rm s}=9$, 
each of which shows finite-size steps 
owing to sublattice systems of $N_{\rm s}=9$. 
Let us return to the main panel of Fig.~\ref{fig3}(c); 
one can recognize that 
the behavior of the large-width steps for $N_{\rm s}=27$ 
comes from the finite-size characteristics of sublattice systems.  
Therefore, the small-width steps originate from 
the interactions of $J_{1}$, which is smaller than $J_{2}$, 
as an effect of perturbation. 
A similar behavior of the large-width steps can clearly be observed 
in the result for $N_{\rm s}=36$ at $m=1/3$, 1/2, and 2/3, 
although the behaviors at $m=1/6$ and 5/6 seem present but weaker. 
At least for $m=1/3$, the large-width steps survive 
for both $N_{\rm s}=27$ and 36. 
In the case of $r=0.3$ in Fig.~\ref{fig3}(b), on the other hand, 
there is a marked difference from the two cases of $r=0.1$ and 2.2; 
one cannot find clear plateaulike behaviors at any height 
including $m=1/3$. 

\begin{figure}[tb]
\begin{center}
  \includegraphics[width=8cm]{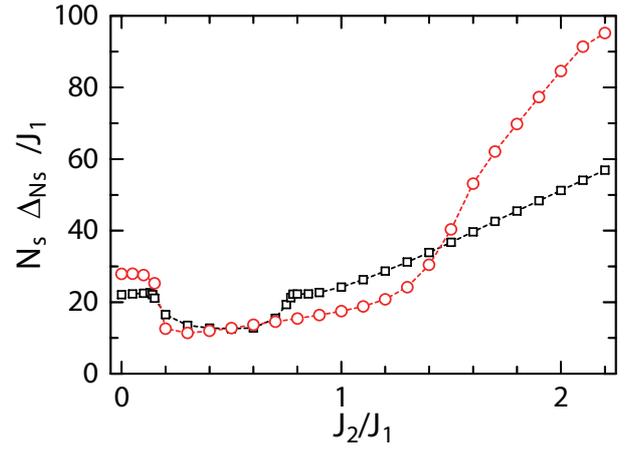}
\end{center}
\caption{
(Color) 
Analysis of the width of the finite-size step at $m=1/3$ 
when the NNN interaction is varied. 
Black squares and red circles denote results for $N_{\rm s}=27$ and 36, 
respectively. 
}
\label{fig4}
\end{figure}

Next, let us investigate the stability of the nonplateau behavior 
at $m=1/3$ in Fig.~\ref{fig3}(b) when $r$ is changed. 
To know the stability,
we start our analysis under the assumption
of a nonplateau situation.
Then, 
the energy per site in the thermodynamic limit $\epsilon(m)$ 
as a function of $m$ is defined as 
\begin{equation}
\frac{E_{r}(N_{\rm s},M)}{N_{\rm s}}\sim \epsilon (m) . 
\end{equation}
If we assume that $\epsilon(m)$ is an analytic function of $m$,
the spin excitation energy would become
\begin{equation}
E_{r}(N_{\rm s},M+1)-E_{r}(N_{\rm s},M)
 \sim\frac{1}{S} 
\left[
\epsilon^{\prime}(m) + \frac{1}{2} \epsilon^{\prime\prime}(m) \frac{1}{N_{\rm s}S}
\right] , 
\end{equation}
which gives the quantity corresponding
to the width of the finite-size step $\Delta_{N_{\rm s}}$ 
at the height of $m$ in the magnetization curve as follows: 
\begin{eqnarray}
\Delta_{N_{\rm s}} &\equiv&
\left[
E_{r}(N_{\rm s},M+1)-E_{r}(N_{\rm s},M) 
\right]
\nonumber \\ 
& & - \left[
E_{r}(N_{\rm s},M)-E_{r}(N_{\rm s},M-1) 
\right]
\nonumber \\
&\sim& \epsilon^{\prime\prime}(m) \frac{1}{N_{\rm s} S^{2}} . 
\label{finite_size_step_width}
\end{eqnarray}
Minimizing the energy of the total Hamiltonian
${\cal H}_{0}+{\cal H}_{\rm Zeeman}$ yields the magnetization
curve at zero temperature using $h=\epsilon^{\prime}(m)/S$.
The field derivative of the magnetization is defined as
$\chi_{\rm mag}\equiv dm/dh = S/\epsilon^{\prime\prime}(m)$. 
When the nonplateau behavior appears,
the derivative should become a nonzero value, namely, $\chi_{\rm mag}\ne 0$. 
In this case, Eq.~(\ref{finite_size_step_width}) can be rewritten as 
\begin{equation}
N_{\rm s} \Delta_{N_{\rm s}} \sim \frac{\epsilon^{\prime\prime}(m)}{S^{2}} . 
\end{equation}
This means that, if there is a region of $r$
where $N_{\rm s} \Delta_{N_{\rm s}}$ is almost constant with increasing $N_{\rm s}$, 
one recognizes that the region is a nonplateau.  
If a plateau exists, on the other hand, 
$N_{\rm s} \Delta_{N_{\rm s}}$ would increase with $N_{\rm s}$. 
Let us apply this argument to the case of $m=1/3$ 
and discuss our present numerical result
of $N_{\rm s} \Delta_{N_{\rm s}}$ shown in Fig.~\ref{fig4},
which is expected to privide some valuable information
concerning whether or not 
the plateau at $m=1/3$ is present. 
From the criteria explained above, 
Fig.~\ref{fig4} shows that the plateau 
is present from $r=0$ to $r\sim 0.15$. 
The behavior markedly changes between $r\sim 0.15$ and $r\sim 0.2$. 
From $r\sim 0.3$ to $r\sim 0.7$, 
one cannot find a significant size dependence 
of $N_{\rm s} \Delta_{N_{\rm s}}$, 
which strongly suggests that the system is in the nonplateau region. 
The presence of this region is a primary result of the present study. 
At $r\sim 0.8$, the behavior of $N_{\rm s} \Delta_{N_{\rm s}}$
for $N_{\rm s}=27$ shows a change in the $r$ dependence; on the other hand, 
a corresponding change is not observed in the result for $N_{\rm s}=36$. 
As a consequence of the change in $N_{\rm s}=27$,
from $r\sim 0.8$ to $r\sim 1.5$,
$N_{\rm s} \Delta_{N_{\rm s}}$ gradually decreases as $N_{\rm s}$ increases; 
the characteristics of this region are unclear at present. 
At $r\sim 1.5$, $N_{\rm s} \Delta_{N_{\rm s}}$ for $N_{\rm s}=36$ shows 
a continuous but marked change in its increase as a function of $r$.  
Above $r\sim 1.6$, therefore, $N_{\rm s} \Delta_{N_{\rm s}}$ 
for a given $r$ clearly increases as $N_{\rm s}$ increases, 
which strongly suggests that the system is in the plateau region. 
The presence of the nonplateau region 
necessarily indicates that
there is a transition point at $r=r_{\rm c1}$
between the point with a plateau at $r=0$ and the nonplateau region. 
The presence of the nonplateau region also 
indicates that 
there is a transition point at $r=r_{\rm c2}$ 
between the point with a plateau in the limit of $r\rightarrow\infty$ 
and the nonplateau region. 

\begin{figure}[tb]
\begin{center}
  \includegraphics[width=8cm]{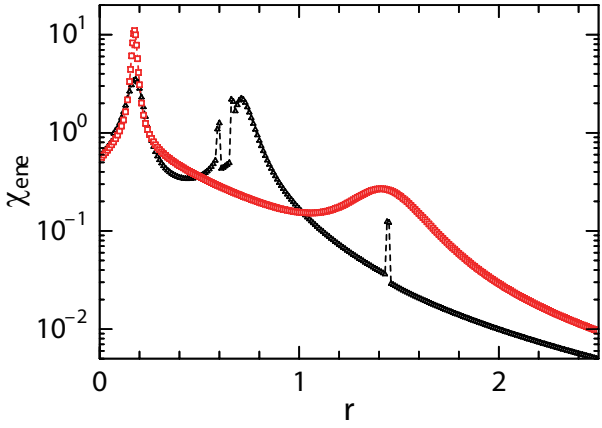}
\end{center}
\caption{
(Color) 
Second derivative of the ground-state energy of the $m=1/3$ state. 
Black triangles and red squares denote results 
for $N_{\rm s}=27$ and 36, respectively. 
}
\label{fig5}
\end{figure}

In order to capture the boundaries $r_{\rm c1}$ and $r_{\rm c2}$ well 
for a given $N_{\rm s}$ when the transition is continuous, 
it is useful to observe 
the second derivative of the ground-state energy 
with respect to the parameter of the model 
of interest\cite{You_second_deri_PRB2010,Albuquerque_second_deri_PRB2010}. 
Such a derivative is defined as 
$
\chi_{\rm ene} \equiv 
- \partial^{2} [E_{r} (N_{\rm s},M)/N_{\rm s}]/\partial r^{2}
$.  
Numerically, we evaluate this quantity using 
\begin{equation}
\chi_{\rm ene} = 
\frac{1}{N_{\rm s}}
\frac{
2E_{r} (N_{\rm s},M)-E_{r+\delta r} (N_{\rm s},M)-E_{r-\delta r} (N_{\rm s},M)
}{(\delta r)^{2}} , 
\label{second_derivative}
\end{equation}
at $M=(1/3)M_{\rm sat}$. We take $\delta r=0.01$. 
The result is depicted in Fig.~\ref{fig5}. 
For $N_{\rm s}=27$, discontinuities are detected 
at $r\sim 0.07$, $\sim 0.60 $, $\sim 0.66$, and $\sim 1.44$. 
For $N_{\rm s}=36$, on the other hand, no discontinuities are observed. 
In spite of the fact that such discontinuities are present
for $N_{\rm s}=27$, two peaks for each $N_{\rm s}$ 
in the continuous $r$ dependence of $\chi_{\rm ene}$ 
are observed 
at $r\sim 0.18$ and $ \sim 0.71$ for $N_{\rm s}=27$ 
and 
at $r\sim 0.17$ and $ \sim 1.41$ for $N_{\rm s}=36$.  
First, let us discuss the behavior of the smaller-$r$ peaks. 
It is notable that 
the position of the smaller-$r$ peaks almost does not change 
with respect to $N_{\rm s}$. 
The position is in good agreement with the position where 
the transition between the small-$r$ plateau region 
and the nonplateau region occurs, which is suggested 
from the examination of $N_{\rm s} \Delta_{N_{\rm s}}$ in Fig.~\ref{fig4}. 
It is reasonable to consider that 
the smaller-$r$ peaks for each $N_{\rm s}$ correspond to the transition. 
Therefore, the occurrence of the transition 
is evident at $r\sim 0.17$, namely, at $r_{\rm c1}\sim 0.17$. 
Next, let us discuss the behavior of the larger-$r$ peaks. 
Although the position of the larger-$r$ peaks shows 
a significantly large change with respect to $N_{\rm s}$, 
both 
$r \sim 0.71$ for $N_{\rm s}=27$ 
and 
$r \sim 1.41$ for $N_{\rm s}=36$ 
are in agreement with the position 
where the behavior of $N_{\rm s} \Delta_{N_{\rm s}}$ 
for each $N_{\rm s}$ markedly changes. 
The observation of the larger-$r$ peaks 
strongly suggests that the transition certainly occurs 
between the nonplateau region and the large-$r$ plateau region,  
although it is difficult to precisely estimate its transition point. 
To summarize, therefore, 
our present analysis results suggest 
that the system shows a plateau at $m=1/3$ for a small $r$, 
that the plateau disappears once at $r=r_{\rm c1}$, and 
that the plateau opens again for $r \simge r_{\rm c2}$.  
In particular, we successfully estimate $r_{\rm c1}\sim 0.17$. 
On the other hand, it is still difficult to estimate $r_{\rm c2}$ precisely. 
Precise estimation of $r_{\rm c2}$
should be carried out in future studies. 

\begin{figure}[tb]
\begin{center}
  \includegraphics[width=8cm]{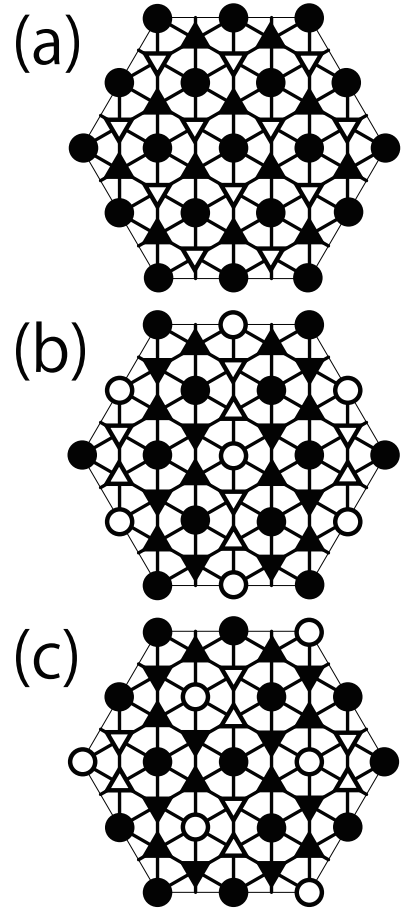}
\end{center}
\caption{
Typical spin configurations characterizing the $m=1/3$ plateau state. 
Each symbol at spin sites (circles, triangles, and reversed triangles)  
corresponds to one of the three sublattices. 
Closed symbols denote that the spin at its site is up; 
on the other hand, 
open symbols denote that the spin at its site is down. 
}
\label{fig6}
\end{figure}

\begin{figure}[tb]
\begin{center}
  \includegraphics[width=8cm]{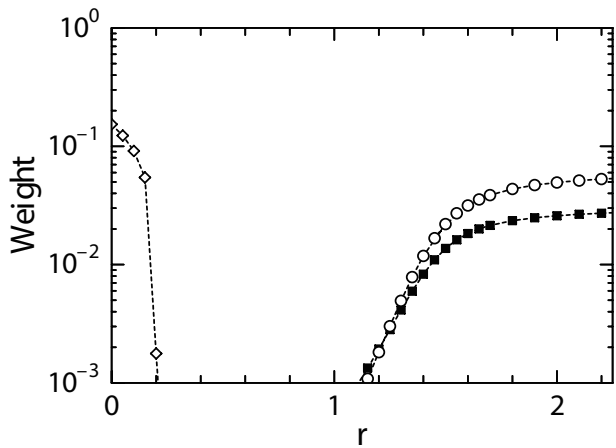}
\end{center}
\caption{
$r$ dependence of weights of typical states in the $m=1/3$ state 
for $N_{\rm s}=36$. 
Open diamonds, closed squares, and open circles 
represent the results corresponding to the states 
illustrated in Figs.~\ref{fig6}(a)-\ref{fig6}(c), 
respectively.
}
\label{fig7}
\end{figure}

The present model of the two limiting cases $J_{2}=0$ and 
$J_{2}\rightarrow\infty$ forms the {\it up-up-down} states at $m=1/3$. 
These states are magnetized under a magnetic field 
and are both considered to be collinear. 
Such collinear states include 
components that are typical, namely, 
they have significantly large weights 
in eigenstates. 
Since we take the basis in our diagonalizations 
so that each element is expressed by either an up spin or a down spin, 
the behaviors of collinear states are captured 
by observing the weights of such elements. 
In the plateau region of small $r$, 
all spins in one subspace among A, B, and C subspaces are down; 
all spins in the remaining two subspaces are up. 
This situation is illustrated in Fig.~\ref{fig6}(a). 
In the plateau region of large $r$, 
on  the other hand,  
in each of the A, B, and C subspaces, 
$(1/9)N_{\rm s}$ spins are down and $(2/9)N_{\rm s}$ spins are up. 
Within each sublattice, 
the positions of $(1/9)N_{\rm s}$ down spins are located 
in the pattern illustrated in Fig.~\ref{fig6}(a).  
Under such situations, 
there are two possible patterns of spin configurations 
of the whole system; the two patterns are illustrated 
in Figs.~\ref{fig6}(b) and \ref{fig6}(c). 
In the configuration in Fig.~\ref{fig6}(b), 
all down spins are located along a linear line. 
In the configuration in Fig.~\ref{fig6}(c), 
on the other hand, 
an island is composed of three down spins; 
each island is located as far from each other as possible. 
Both patterns can produce a plateau at $m=1/3$.  
Let us then examine the weights of the three patterns. 
The weights are evaluated 
as the sum of squared coefficients of spin configurations 
in the normalized ground state at $m=1/3$ 
when the selected configurations are linked 
to a pattern illustrated in Figs.~\ref{fig6}(a)-\ref{fig6}(c) 
by a translational or rotational symmetry of the system.  
Let us observe the weights for the $N_{\rm s}=36$ system 
that show no discontinuities in the results presented in Fig.~\ref{fig5};  
the results are depicted in Fig.~\ref{fig7}. 
It is observed that the pattern in Fig.~\ref{fig6}(a) 
has a large weight for $r$ that is smaller than $r\sim 0.15$. 
At $r\sim 0.2$ and above, this pattern loses its weight. 
This behavior is in good agreement with the behavior in Fig.~\ref{fig4} 
and $r_{\rm c1}\sim 0.17$ obtained from Fig.~\ref{fig5}. 
For $r > 1$, on the other hand, 
the weights gradually increase  
for the patterns in Figs.~\ref{fig6}(b) and \ref{fig6}(c). 
At $r\sim 1.4$ and above, 
both weights are significantly large; 
the weight for the pattern in Fig.~\ref{fig6}(b) 
is larger than that in Fig.~\ref{fig6}(c). 
This behavior suggests that 
the $m=1/3$ plateau in the large-$r$ region 
originates from the formation of the spin configuration 
of the pattern in Fig.~\ref{fig6}(b).   
This suggestion should be confirmed by other calculations 
in future studies. 

In the present system, 
the above-mentioned states are magnetized under a magnetic field 
and are 
both considered to be collinear;  
the positions of up or down spins 
are different between the two cases,  
as described in the last paragraph. 
The present study reveals that 
an intermediate region without a plateau 
appears between the two different {\it up-up-down} states. 
A similar situation is known in the square-lattice Heisenberg 
antiferromagnet with NNN interactions, the so-called $J_{1}$-$J_{2}$ 
model\cite{Dagotto1989,Nishimori1990,Poilblanc1991,Chubukov1991,
Zhitomirsky1996,Singh1999,Capriotti2000,Capriotti2001,Li2012,
Jiang2012,Hu2013,Gong2014,Richter2015,Morita_JPSJ2015,Wang2016}. 
The properties of this system have long been investigated 
between the two phases, one is the ordinary N$\acute{\rm e}$el phase 
and the other is a phase of another collinear state 
when $J_{2}/J_{1}$ is varied. 
A recent variational Monte-Carlo study\cite{Morita_JPSJ2015} suggested 
that such an intermediate region is divided into more than one phase. 
From an analogy of the square-lattice $J_{1}$-$J_{2}$ model,  
the obtained intermediate region in the present model (\ref{Hamiltonian}) 
may be composed of complex phases. 
This issue should be investigated in future studies. 

At $m=1/3$, the magnetization plateau 
appears in various frustrated systems.
In the kagome-lattice antiferromagnet, the plateau 
is accompanied by anomalous critical behavior 
with critical exponents different from the exponent $\delta=1$, 
which is typical of two-dimensional systems, 
just outside of the edges of the $m=1/3$ state 
in the magnetization curve\cite{HN_TSakai_kgm_1_3,HN_TSakai_ramp_2010}, 
where the critical exponent $\delta$ is defined as 
$|m-m_{\rm c}|\sim |h-h_{\rm c}|^{1/\delta}$ 
near the transition point $h_{\rm c}$. 
The Cairo-pentagon-lattice antiferromagnet\cite{HNakano_Cairo_lt,
Isoda_Cairo_full} and
square-kagome lattice antiferromagnet\cite{shuriken_lett,shuriken_dist} 
reveal a magnetization jump at an edge of the $m=1/3$ state. 
A similar jump also appears for the kagome-lattice antiferromagnet 
with a distortion\cite{Hida2001,HN_kgm_dist}. 
On the other hand, the magnetization plateau
of the triangular-lattice antiferromagnet without NNN interactions 
shows the typical exponent $\delta=1$ 
in the two-dimensional systems\cite{Sakai_HN_PRBR}.  
The critical behavior around the edges of the $m=1/3$ state 
of the present system should be clarified in future studies 
in which NNN interactions are switched on.

\section{Conclusion} 

We investigated the ground-state magnetization process 
of the spin-$1/2$ Heisenberg antiferromagnet on a triangular lattice 
with next-nearest-neighbor interactions 
by the numerical-diagonalization method. 
For small amplitudes of NNN interactions, 
the $m=1/3$ plateau of this system survives; 
a plateau of the same height exists 
for large amplitudes of NNN interactions. 
We have found that, in an intermediate region, 
this plateau disappears. 
In particular, 
the boundary of the intermediate region on the smaller-$J_{2}$ side 
is found to be $J_{2}\sim 0.17 J_{1}$, 
which should be examined to obtain a precise estimate in the future. 
To precisely know where the boundary on the larger-$J_{2}$ side is, 
investigations of larger systems are required. 
Further study 
of phenomena due to NNN interactions 
in a triangular-lattice antiferromagnet 
would greatly contribute 
to our understanding of the frustration effect 
in quantum spin systems.

\section*{Acknowledgments}
We wish to thank 
Professor 
H.~Tanaka 
and Professor N. Todoroki 
for fruitful discussions.
This work was partly supported 
by JSPS KAKENHI Grant Numbers 
16K05418, 16K05419, and 16H01080 (JPhysics). 
Nonhybrid thread-parallel calculations
in numerical diagonalizations were based on TITPACK version 2
coded by H. Nishimori. 
In this research, we used the computational resources of the K computer 
provided by the RIKEN Advanced Institute for Computational Science 
through the HPCI System Research projects 
(Project ID: hp170017, hp170028, hp170070, and hp170207). 
Some of the computations were 
performed using facilities of 
the Department of Simulation Science, 
National Institute for Fusion Science; 
Institute for Solid State Physics, The University of Tokyo;  
and Supercomputing Division, 
Information Technology Center, The University of Tokyo. 
This work was partly supported 
by the Strategic Programs for Innovative Research; 
the Ministry of Education, Culture, Sports, Science 
and Technology of Japan; 
and the Computational Materials Science Initiative, Japan. 



\begin{thebibliography}{99} 
\bibitem{Anderson_tri} 
P.~W.~Anderson, Mater. Res. Bull. \textbf{8}, 153 (1973).
\bibitem{Huse_Elser} 
D.~A.~Huse and V.~Elser, Phys. Rev. Lett. \textbf{60}, 2531 (1988).
\bibitem{Jolicour_LGuillou}
Th.~Jolicour and J.~C.~Le~Guillou, Phys. Rev. B \textbf{40}, 2727 (1989).
\bibitem{Singh_Huse}
R.~R.~P.~Singh and D.~A.~Huse, Phys. Rev. Lett. \textbf{68}, 1766 (1992).
\bibitem{Bernu1992}
\label{Bernu1992}
B.~Bernu, C.~Lhuillier, and L.~Pierre, 
Phys.~Rev.~Lett. \textbf{69}, 2590 (1992).
\bibitem{Cubukov1992}
\label{Cubukov1992}
A.~V.~Chubukov and Th.~Jolicoeur, 
Phys.~Rev.~B. \textbf{46}, 11137 (1992).
\bibitem{Bernu1994}
B.~Bernu, P.~Lecheminant, C.~Lhuillier, and L.~Pierre,
Phys. Rev. B \textbf{50}, 10048 (1994).
\bibitem{Leung_Runge}
P.~W.~Leung and K.~J.~Runge, Phys. Rev. B \textbf{47}, 5861 (1993).
\bibitem{Lecheminant1995}
\label{Lecheminant1995} 
P.~Lecheminant, B.~Bernu, C.~Lhuillier, and L.~Pierre, 
Phys.~Rev.~B \textbf{52}, 6647 (1995).
\bibitem{Richter_lecture2004}
 J.~Richter, J.~Schulenburg, and A.~Honecker,
{\it Lecture Notes in Physics} (Springer, Heidelberg, 2004) Vol. 645, p. 85.
\bibitem{Sakai_HN_PRBR}
\label{Sakai_HN_PRBR}
T.~Sakai and H.~Nakano, Phys.~Rev.~B \textbf{83}, 100405(R) (2011). 
\bibitem{DYamamoto2013}
D.~Yamamoto, G.~Marmorini, and I.~Danshita, 
Phys. Rev. Lett. \textbf{112}, 127203 (2014)
\bibitem{Weihong1999}
Z.~Weihong, R.~H.~McKenzie, and R.~R.~P.~Singh,
Phys. Rev. B \textbf{59}, 14367 (1999).
\bibitem{Yunoki2006}
S.~Yunoki and S.~Sollera, Phys. Rev. B \textbf{74}, 014408 (2006).
\bibitem{Starykh2007}
O.~A.~Starykh and L.~Balents, Phys. Rev. Lett. \textbf{98}, 077205 (2007).
\bibitem{Heidarian2009}
D.~Heidarian, S.~Sollera, and F.~Becca,
Phys. Rev. B \textbf{80}, 012404 (2009).
\bibitem{Reuther2011}
J.~Reuther and R.~Thomale, Phys. Rev. B \textbf{83}, 024402 (2011).
\bibitem{Weichselbaum2011}
A.~Weichselbaum and S.~R.~White, Phys. Rev. B \textbf{84}, 245130 (2011).
\bibitem{Ghamari2011}
S.~Ghamari, C.~Kallin, S.~S.~Lee, and E.~S.~Sorensen,
Phys. Rev. B \textbf{84}, 174415 (2011).
\bibitem{Harada2011}
K.~Harada, Phys. Rev. B \textbf{86}, 184421 (2012).
\bibitem{s1tri_LRO}
\label{s1tri_LRO}
H.~Nakano, S.~Todo, and T.~Sakai, 
J.~Phys.~Soc.~Jpn. \textbf{82}, 043715 (2013).
\bibitem{Starykh_review2015}
O.~A.~Starykh, Rep. Prog. Phys. \textbf{78}, 052502 (2015).
\bibitem{Kulagin_PRL2013}
S.~A.~Kulagin, N.~Prokof'ev, O.~A.~Starykh, B.~Svistunov, and C.~N.~Varney, 
Phys. Rev. Lett. \textbf{110}, 070601 (2013).
\bibitem{Alicea_PRL2009}
J.~Alicea, A.~V.~Chubukov, and O.~A.~Starykh, 
Phys. Rev. Lett. \textbf{102}, 137201 (2009). 
\bibitem{Shirata_Shalf_PRL2012}
Y.~Shirata, H.~Tanaka, A.~Matsuo, and K.~Kindo, 
Phys. Rev. Lett. \textbf{108}, 057205 (2012).   
\bibitem{Shirata_S1tri_2011}
Y. Shirata, H. Tanaka, T. Ono, A. Matsuo, K. Kindo, and H. Nakano, 
J. Phys. Soc. Jpn. \textbf{80} (2011) 093702.
\bibitem{KWatanabe_JPSJ2015}
K.~Watanabe, H.~Kawamura, H.~Nakano, and T.~Sakai,
J.~Phys.~Soc.~Jpn. \textbf{83}, 034714 (2014).
\bibitem{Shimokawa_JPSJ2015}
T.~Shimokawa, K.~Watanabe, and H.~Kawamura,
Phys.~Rev.~B \textbf{92}, 134407 (2015). 
\bibitem{Wietek_arXiv2016}
\label{Wietek_arXiv2016}
A.~Wietek and A.~M.~L$\ddot{\rm a}$uchli, arXiv: 1604.07829. 
\bibitem{Bieri_PRL}
R.~V.~Mishmash, J.~R.~Garrison, S.~Bieri, and C.~Xu,
Phys.~Rev.~Lett. \textbf{111}, 157203 (2013). 
\bibitem{HN_Terai}
\label{HN_Terai}
H.~Nakano and A.~Terai, 
J.~Phys.~Soc.~Jpn. \textbf{78}, 014003 (2009).
\bibitem{kgm_gap}
\label{kgm_gap}
H.~Nakano and T.~Sakai, 
J.~Phys.~Soc.~Jpn. \textbf{80}, 053704 (2011).
\bibitem{HN_TSakai_kgm_1_3}
\label{HN_TSakai_kgm_1_3}
H.~Nakano and T.~Sakai, 
J.~Phys.~Soc.~Jpn. \textbf{83}, 104710 (2014).
\bibitem{HN_TSakai_kgm_S}
H.~Nakano and T.~Sakai, 
J.~Phys.~Soc.~Jpn. \textbf{84}, 063705 (2015).
\bibitem{HN_YHasegawa_TSakai_dist_shuriken}
H.~Nakano, Y.~Hasegawa, and T.~Sakai,
J.~Phys.~Soc.~Jpn. \textbf{84}, 114703 (2015). 
\bibitem{You_second_deri_PRB2010}
W.-L.~You, Y.-W.~Li, and S.-J.~Gu, 
Phys. Rev. E \textbf{76}, 022101 (2007). 
 A. F. Albuquerque, F. Alet, C. Sire, and S. Capponi, 
Phys. Rev. B \textbf{81}, 064418 (2010).
\bibitem{Albuquerque_second_deri_PRB2010}
A.~F.~Albuquerque, F.~Alet, C.~Sire, and S.~Capponi, 
Phys. Rev. B \textbf{81}, 064418 (2010).
\bibitem{Dagotto1989}
E.~Dagotto and A.~Moreo, 
Phys.~Rev.~Lett. \textbf{63}, 2148 (1989). 
\bibitem{Nishimori1990}
H.~Nishimori and Y.~Saika, 
J.~Phys.~Soc.~Jpn. \textbf{59}, 4454 (1990).
\bibitem{Poilblanc1991}
D.~Poilblanc, E.~Gagliano, S.~Bacci, and E.~Dagotto, 
Phys.~Rev.~B \textbf{43}, 10970 (1991).
\bibitem{Chubukov1991}
A.~V.~Chubukov and T.~Jolicoeur, 
Phys.~Rev.~B \textbf{44}, 12050 (1991).
\bibitem{Zhitomirsky1996}
M.~E.~Zhitomirsky and K.~Ueda, 
Phys.~Rev.~B \textbf{54}, 9007 (1996).
\bibitem{Singh1999}
R.~R.~P.~Singh, Z.~Weihong, C.~J.~Hamer, and J.~Oitmaa, 
Phys.~Rev.~B \textbf{60}, 7278 (1999). 
\bibitem{Capriotti2000}
L.~Capriotti and S.~Sorella, 
Phys.~Rev.~Lett. \textbf{84}, 3173 (2000).
\bibitem{Capriotti2001}
L.~Capriotti, F.~Becca, A.~Parola, and S.~Sorella, 
Phys.~Rev.~Lett. \textbf{87}, 097201 (2001).
\bibitem{Li2012}
T.~Li, F.~Becca, W.~Hu, and S.~Sorella, 
Phys.~Rev.~B \textbf{86}, 075111 (2012).
\bibitem{Jiang2012}
H.-C.~Jiang, H.~Yao, and L.~Balents, 
Phys.~Rev.~B \textbf{86}, 024424 (2012).
\bibitem{Hu2013}
W.-J.~Hu, F.~Becca, A.~Parola, and S.~Sorella, 
Phys.~Rev.~B \textbf{88}, 060402 (2013).
\bibitem{Gong2014}
S.-S.~Gong, W.~Zhu, D.~N.~Sheng, O.~I.~Motrunich, and M.~P.~A.~Fisher,
Phys.~Rev.~Lett. \textbf{113}, 027201 (2014).
\bibitem{Richter2015}
J.~Richter, R.~Zinke, and D.~Farnell, 
Eur.~Phys.~J. B \textbf{88}, 2 (2015). 
\bibitem{Morita_JPSJ2015}
\label{Morita_JPSJ2015}
S.~Morita, R.~Kaneko, and M.~Imada, 
J.~Phys.~Soc.~Jpn. \textbf{84}, 024720 (2015).
\bibitem{Wang2016}
L.~Wang, Z.-C.~Gu, F.~Verstraete, and X.-G.~Wen, 
Phys.~Rev.~B \textbf{94}, 075143 (2016). 
%
%
\bibitem{HN_TSakai_ramp_2010}
\label{HN_TSakai_ramp_2010}
H.~Nakano and T.~Sakai, J.~Phys.~Soc.~Jpn. \textbf{79}, 053707 (2010). 
\bibitem{HNakano_Cairo_lt}
\label{HNakano_Cairo_lt}
H.~Nakano, M.~Isoda, and T.~Sakai, 
J.~Phys.~Soc.~Jpn. \textbf{83}, 053702 (2014). 
\bibitem{Isoda_Cairo_full}
\label{Isoda_Cairo_full}
M.~Isoda, H.~Nakano, and T.~Sakai, 
J.~Phys.~Soc.~Jpn. \textbf{83}, 084710 (2014). 
\bibitem{shuriken_lett}
\label{shuriken_lett}
H.~Nakano and T.~Sakai, 
J.~Phys.~Soc.~Jpn. \textbf{82}, 083709 (2013).
\bibitem{shuriken_dist}
\label{shuriken_dist}
H.~Nakano, Y.~Hasegawa, and T.~Sakai, 
J.~Phys.~Soc.~Jpn. \textbf{84}, 114703 (2015).
\bibitem{Hida2001}
\label{Hida2001}
K.~Hida, J.~Phys.~Soc.~Jpn. \textbf{70}, 3673 (2001).
\bibitem{HN_kgm_dist}
\label{HN_kgm_dist}
H.~Nakano, T.~Sakai, and Y.~Hasegawa,
J.~Phys.~Soc.~Jpn. \textbf{83}, 084709 (2014).
%
%
%
%
%
%
\end{thebibliography}
\end{document}